%% file: charge_sensitivity_nanolett.tex
\documentclass[journal=nalefd,manuscript=letter,layout=twocolumn]{achemso}

\usepackage{array}
\usepackage{booktabs}
\usepackage{listings}
\usepackage{lmodern}
\usepackage{mathpazo}
\usepackage{microtype}
\usepackage{caption}
\usepackage{float}
\usepackage{geometry}
\usepackage{natbib}
\usepackage{setspace}
\usepackage{xkeyval}

\usepackage{subfigure} %subfigures
\usepackage{textcomp} %\textmu
\usepackage{upgreek} %nonitalic mu in math environment
\usepackage{amssymb}
\usepackage{color}
\usepackage{titlesec}
\titleformat{\section}[runin]
  {\normalfont\large\bfseries}{\thesection}{}{}
\titleformat{\subsection}[runin]
  {\normalfont\normalsize\bfseries}{\thesubsection}{}{}
\newcommand{\subscript}[1]{\ensuremath{_{\textrm{#1}}}}
\definecolor{blue}{rgb}{0.0,0.0,1}
\definecolor{AI}{rgb}{1.0,0.0,0}

\definecolor{PTH}{rgb}{0.0,0.5,1.0}

\definecolor{revision}{rgb}{0.0,0.0,1}

\author{Pasi H\"akkinen}
\affiliation[Aalto University]
{Low Temperature Laboratory, O.V. Lounasmaa Laboratory, Aalto University, FI-00076 AALTO, Finland}

\author{Andreas Isacsson}
\affiliation[Chalmers University]
{Department of Applied Physics, Chalmers University of Technology, SE-412 96 G\"oteborg, Sweden}

\author{Alexander Savin}
\affiliation[Aalto University]
{Low Temperature Laboratory, O.V. Lounasmaa Laboratory, Aalto University, FI-00076 AALTO, Finland}

\author{Jaakko Sulkko}
\affiliation[Aalto University]
{Low Temperature Laboratory, O.V. Lounasmaa Laboratory, Aalto University, FI-00076 AALTO, Finland}

\author{Pertti Hakonen}
\email{pertti.hakonen@aalto.fi}
\affiliation[Aalto University]
{Low Temperature Laboratory, O.V. Lounasmaa Laboratory, Aalto University, FI-00076 AALTO, Finland}

\title[]{Charge sensitivity enhancement via mechanical oscillation in suspended carbon nanotube devices}

\abbreviations{IR,NMR,UV}
\keywords{Electrometry, charge sensing, single-walled carbon nanotube, nanomechanical resonator}

\begin{document}

%\begin{tocentry}
%%Mechanical-oscillation-induced change in the conductance of a suspended single-walled carbon nanotube, together with the gate voltage dependence of the resonance frequency, can increase the transconductance of the device through \(\partial G/\partial\left(\delta f\right)\times\partial f_{0}/\partial V_{\rm G}\). This enhancement enables reaching record charge sensitivity of single-electron transistors. The best result was reached by biasing the resonator to a point where the single-electron-tunneling induced Duffing constant is canceled out, and the mechanical properties are determined mainly by the \(5^{\rm th}\) order conservative nonlinearity. This not only gives the optimal sensitivity of 0.97 \textmu e/\(\sqrt\mathrm{Hz}\) at \(\sim1.3\) kHz but also results in an interesting mechanical response.
%\includegraphics{toc_graphics}
%\end{tocentry}

\begin{abstract}
\emph{Single electron transistors (SETs) fabricated from single-walled carbon nanotubes (SWNTs) can be operated as highly sensitive charge detectors reaching sensitivity levels comparable to metallic radio frequency SETs (rf-SETs). Here we demonstrate how the charge sensitivity of the device can be improved by using the mechanical oscillations of a single-walled carbon nanotube quantum dot. To optimize the charge sensitivity $\delta Q$, we drive the mechanical resonator far into the nonlinear regime and bias it to an operating point where the mechanical third order nonlinearity is cancelled out. This way we enhance $\delta Q$, from 6 \textmu e/\(\sqrt\mathrm{Hz}\) for the static case, to 0.97 \textmu e/$\sqrt\mathrm{Hz}$, at a probe frequency of \(\sim1.3\) kHz.}
\end{abstract}


\subsection*{}
Nanoelectromechanical resonators are often praised
for their high sensitivity for detection of mass, force and
charge. In particular carbon nanotube (CNT) resonators have on several occasions proven to yield unprecedented sensitivities for both
mass~\cite{Jensen2008,Chaste2012,Bockrath2008}, and
force~\cite{Moser2013}.

However, for measuring charge, the Single
Electron Transistor (SET)~\cite{Likharev1999} remains the record
holder in electrometer sensitivity.~\cite{Brenning2006}. With an
operating principle based on Coulomb blockade, its conductance is very sensitive to the island charge $Q$, and hence, the gate voltage $V_{\rm G}$. Gate voltage changes, corresponding to sub-electron variations of $Q$, cause a measurable change in source-drain current \(I_{\rm SD}\) through modulating the transconductance $G_{\rm m}=\partial I_{\rm SD}/\partial V_{\rm G}$. This allows for charge sensitivities of the order $\delta Q\approx 1\ \upmu e/\sqrt{\mathrm{Hz}}$.\cite{Devoret2000}

Although a  CNT-SET can demonstrate good charge sensitivity in itself~\cite{Roschier2001,Andresen2008}, we show here that driving it into mechanical resonance can be used for further enhancement. The improvement is enabled by the very strong coupling between charge dynamics and mechanical motion observed in suspended single-wall carbon nanotube SETs (SWNT-SETs).~\cite{Steele2009b,Lassagne2009,Meerwaldt2012,Benyamini2014} The strong coupling causes mechanically induced changes to the transconductance which are very sensitive to the gate charge.

To achieve high charge sensitivity, noise must be minimized. For instance in a
SET, low frequency \(1/f^{\alpha}\)-noise is usually circumvented by
measuring reflected power in the radio-frequency range (rf-SET)
instead of direct current.~\cite{Schoelkopf1998}.  While a strong
coupling between charge and mechanical motion is necessary for high
sensitivity in the SWNT-SET resonator setup, it is accompanied by
strong induced mechanical nonlinearities. The interplay between noise
and nonlinear phenomena in nanoresonators is currently an active
research topic~\cite{Villanueva2013, Eichler2013} and it is well known that
nonlinearities, in general, limit the dynamic range of nanomechanical
sensing devices~\cite{Postma2005}.

\begin{figure}[t]
  \begin{center}
      \includegraphics[width=\linewidth]{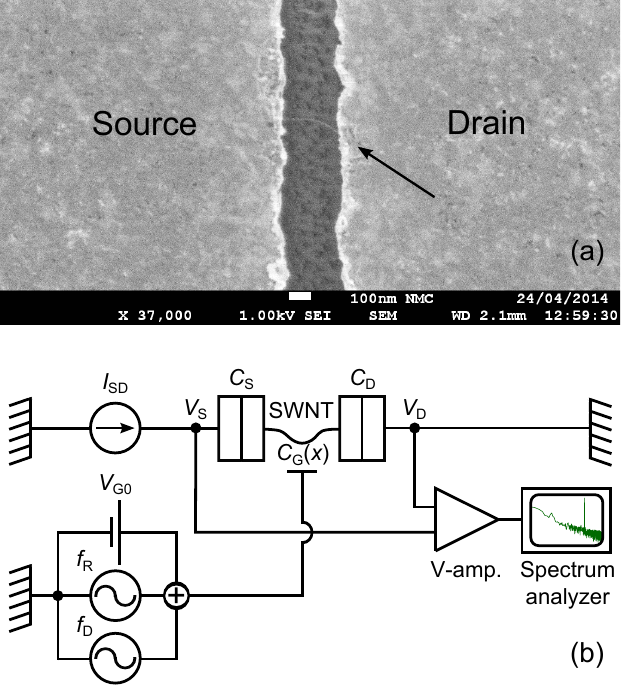}\\
\end{center}
\caption{Single-Walled Carbon-Nanotube Single-Electron-Transistor (SWNT-SET) mechanical resonator. (a) Scanning electron microscope image of device with a suspended CNT between source and drain electrodes. (b) Measurement setup. Mechanical motion is induced by driving the gate with a frequency $f_{\rm D}$. The current-voltage characteristics at resonance is probed at a readout frequency $f_{\rm R}\ll f_{\rm D}$. Using a spectrum analyzer, the charge sensitivity $\delta Q$ is determined from the signal-to-noise ratio at the readout frequency $f_{\rm R}$.
	\label{fig:sysfig}}
\end{figure}

Here we present measurements on a SWNT-SET resonator with strong
electromechanical coupling. The observed resonances fit well to a
model containing gate-tunable 3$^{\rm rd}$ and 5$^{\rm th}$ order electrically induced conservative nonlinearities. We find an optimal enhancement of the charge sensitivity when the resonator is tuned to a bias point where the 3$^{\rm rd}$ order nonlinearity is cancelled out. The obtained charge sensitivity on resonance, $\delta Q_{\rm res.}=0.97\)~\textmu\(e/\sqrt{\mathrm{Hz}}$, should be compared to the sensitivity at static device operation, $\delta Q_{\rm stat.}=6\)~\textmu\(e/\sqrt{\mathrm{Hz}}$ at \(\sim 1.3\) kHz.

\subsection*{Device fabrication and characterization.} In our devices, SWNTs are placed on top of pre-fabricated Nb/Pt electrodes as a final step of the sample fabrication process using chemical vapor deposition growth.~\cite{Kong1998a} This enables production of ultra-clean carbon nanotube resonators with excellent electrical and mechanical properties. Substrate preparation starts by sputtering 200~nm of Nb and 20~nm of Pt on highly doped Si wafer that is capped with 267~nm SiO\(_2\). Next, an aluminium etch mask is fabricated with standard electron beam lithography methods and e-beam evaporation. Because platinum is relatively inert against reactive ion etching (RIE) it is ion beam milled prior to the RIE step that removes the Nb layer. The remaining Al mask is finally etched by immersing the sample in an aqueous 20~\% H\subscript{3}PO\subscript{4} bath.

Preselection of devices is done at room temperature and at cryogenic conditions, where samples demonstrating characteristic transport properties of a single SWNT are cooled to 50 mK in a dilution cryostat. All data reported in this paper are measured at the base temperature of the cryostat. The source and drain electrodes are wirebonded in a four probe configuration to low frequency wiring of the cryostat. Each line contains a segment of Thermocoax cable for attenuating high frequency electromagnetic noise and a three-stage RC filter with 25~kHz cut-off frequency. The doped silicon substrate that serves as a back gate is attached to a line having only high frequency filtering, which enables actuation of mechanical motion up to 600~MHz.

One sample was measured in two separate cool downs, both before and after current annealing. Similar results were consistently obtained even though sample properties changed significantly: mechanical characteristics due to current annealing and electrical contacts due to thermal cycling. The device shows clear Coulomb blockade behavior at cryogenic temperatures. Fit to the simple model given in the supplementary material Section S1 yields the values for the tunneling rates, capacitances and electron temperature of the device.~\cite{Averin1991} These parameters were found to be $C_{\rm S}=3.4$~aF, $C_{\rm D}=7.2$~aF, and $C_{\rm G}=0.34$~aF for the source, drain, and gate capacitances, respectively, in the vicinity of the degeneracy point $V_{\rm G}=V_{\rm deg}\approx -11.5$~V used for the charge sensing experiment described below. In addition, a stray capacitance $C_0=1.0$~aF contributes to the total capacitance $C_{\Sigma}=C_{\rm S}+C_{\rm D}+C_{\rm G}+C_0$. Fitted effective electronic temperature $T^*\approx 4.5$~K is high, because the model neglects the lifetime broadening in strong tunneling regime.

Mechanical motion of the resonator is actuated by applying a static
dc-bias to the gate with a superimposed ac-field with drive frequency
$f_{\rm D}$, i.e. $V_{\rm G}=V_{{\rm G}0}+\delta V(t)$.  Considering
only very small vibrations around the resonator equilibrium position
(amplitudes $\ll 1$~nm), the equation of motion for the
fundamental flexural mode coordinate $x$ is
\begin{eqnarray}
&&\ddot{x}+\gamma_0\dot{x}+\Omega_0^2 x=\frac{1}{2m}C_{\rm G}'(x) V_{\rm
  G}^2\nonumber\\
&&\times\left(1-2\frac{\left<Q(t)\right>+Q^*+V_{\rm G}C_{\rm G}(x)}{V_{\rm
    G}C_\Sigma}\right).
\label{eq:eom1}
\end{eqnarray}
Here $\Omega_0$ is the resonant frequency at zero gate bias, $\Omega_0/\gamma_0$ the bare quality factor of the resonator, $m$ the effective mode mass, and \(\left<Q\left(t\right)\right>\) the average charge of the CNT island. Charge $Q^*=V_{\rm S}C_{\rm S}+V_{\rm D}C_{\rm D}$ is assumed to be independent of resonator deflection $x$. Here \(V_{\rm S}\) and \(V_{\rm D}\) are source and drain voltages, respectively. We have neglected terms of second order in the dot potential in the force on the right hand side of Equation~\ref{eq:eom1}, as $|\left[\left<Q(t)\right>+Q^*+V_{\rm G}C_{\rm G}(x)\right]/{V_{\rm G}C_\Sigma}|\ll 1$ in the vicinity of the degeneracy point.

The motion is detected using the rectification technique~\cite{Steele2009b} which relies on the nonlinear conductance characteristics of the SWNT QD. In the Coulomb blockade regime, the conductance \(G\) of the QD is determined by its average charge. This charge can be changed by varying the product of gate voltage \(V_{\rm G}\) and gate capacitance \(C_{\rm G}\). The corresponding source-drain current is thus to a good approximation a function of the product $V_{\rm G}C_{\rm G}$. Hence, if the current in the static regime is given by $I(t)=I(V_{\rm G}(t))$, the average dc-current in the mechanical resonance regime is found from the time average
\begin{equation}
I_{\rm SD}\equiv\left<I\right>=\frac{1}{T}\int_0^T dt\, I(V_{\rm G}(t)C_{\rm
  G}(x(t))/C_{\rm G}(0)).
\label{eq:Iav}
\end{equation}

In the limit of small vibration amplitudes, expanding the above expression in $x$ gives a mechanical contribution to the
dc-current $\propto \left<{x^2}\right>$. Therefore, by applying voltage bias and measuring the output current while sweeping the drive frequency across the mechanical resonance frequency, we record a line shape proportional to squared oscillation amplitude. For larger amplitudes the gate induced charge traverses the entire Coulomb peak, see Section S2. Using the full integral expression in Equation~\ref{eq:Iav} is then necessary for finding the current.

\begin{figure}[htbp]
  \begin{center}
      \includegraphics[width=\linewidth]{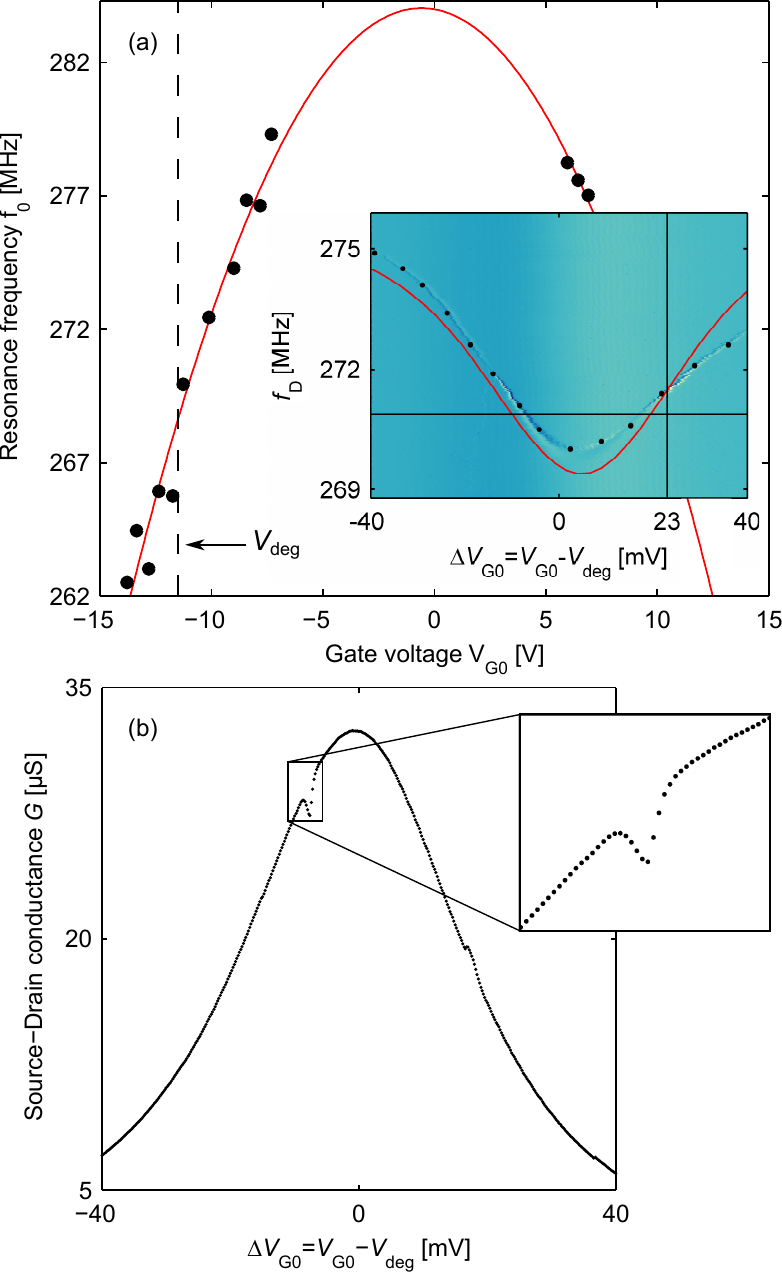}
    \caption{ (a) Measurement of resonance frequency \(f_0\) versus      gate bias \(V_{{\rm G}0}\) verifies that the acquired resonance signal is of mechanical origin. The electrostatic softening allows the resonance frequency to be tuned down from approximately 280 MHz by 18 MHz within the presented \(V_{{\rm G}0}\) range. The inset plots the transconductance of the device against drive frequency and gate voltage. A signal from the mechanical resonance is seen as a faint curve imposed on to the Coulomb effect induced variation. The measurement is made around a charge degeneracy that is located at \(V_{{\rm G}0}=V_{\rm deg}\approx-11.5\)~V using \(V_{\rm SD}=0.25\)~mV. (b) Conductance along the horizontal line of the inset in (a) shows the mechanical-oscillation-induced conductance variation on both sides of the Coulomb peak. The inset reveals that the slope \(\partial G/\partial V_{{\rm G}0}\) can be significantly increased by the mechanical motion.}
     \label{fig_f0_vs_Vg}
 \end{center}
\end{figure}

Figure~\ref{fig_f0_vs_Vg}(a) shows the behavior of the resonance frequency
\(f_{0}=\omega_0/2\pi=(\Omega_0+\Delta\omega_0)/2\pi\) as a function
of gate dc-bias \(V_{{\rm G}0}\). To relate the frequency tuning to model parameters, we expand the right hand side of Equation~\ref{eq:eom1} to lowest order in the
displacement $x$ and identify the coefficients. Keeping the dominant tuning
contributions yields
\begin{eqnarray}
&&\Delta\omega_0=\omega_0-\Omega_0\approx-\frac{C_{\rm G}''V_{\rm G}^2}{4m\Omega_0}\nonumber\\
&&+\frac{V_{\rm G}^2C_{\rm G}'^2}{2m\Omega_0C_{\rm G}C_\Sigma}\frac{\partial \left<Q(t)\right>}{\partial V_{\rm G}}.
\label{eq:tune1}
\end{eqnarray}
The first term in the equation gives the global frequency profile seen in Figure~\ref{fig_f0_vs_Vg}(a) while the second term gives rise to dynamical softening in the vicinity of the degeneracy
points. Fitting the global downward parabolic shape arising from electrostatic softening gives ${C_{\rm G}''}/{2m}\approx 2.83\times 10^{15}$~(Vs)$^{-2}$.

The dynamical softening at \(V_{\rm G0}=V_{\rm deg}\approx-11.5\) V is visible in the inset of Figure~\ref{fig_f0_vs_Vg}(a).  It shows a surface plot of the transconductance of the device when the gate bias is swept across a charge degeneracy point using different drive frequencies \(f_{\rm D}\). The mechanically induced transconductance change is seen as the slowly varying curve imposed on to the variation arising from Coulomb effect. It shows softening behavior according to Equation~\ref{eq:tune1}. Fitting the dynamic softening at the degeneracy point gives $C_{\rm G}'^2/m\approx 1.4\times 10^{-3}$~F$^2/$m$^{2}$kg. Separate fitting parameters acquired from global and local resonance frequency tunings yield effective mode mass \(m=1.7\times10^{-21}\) kg when a capacitance model of wire on infinite plate is used.

The source-drain conductance is plotted in Figure~\ref{fig_f0_vs_Vg}(b) along the black horizontal line in the inset to Figure~\ref{fig_f0_vs_Vg}(a). The mechanical-oscillation-induced local conductance change is highlighted in the inset. It demonstrates that the local slope of the conductance $\partial G/\partial V_{\rm G0}$ can be significantly increased under resonant conditions and, hence, the device becomes more sensitive to gate charge variation in that region.

To understand the origin of this enhancement, consider a linear
resonator where only the resonant frequency $f_0$ depends on
gate voltage. Defining the detuning \(\delta f=f_{\rm D}-f_{0}\), the
conductance $G$ can then be written:
\begin{equation}
G\left(V_{{\rm
    G}0},\delta f(V_{{\rm G}0})\right)=G\left(V_{{\rm G}0},f_{\rm D}-f_{0}\left(V_{{\rm G}0}\right)\right).
\label{eq:Grel}
\end{equation}
This gives a vibrational contribution to the slope
\begin{equation}
\frac{dG}{dV_{\rm G}}=\frac{\partial G}{\partial V_{\rm G}}-\frac{\partial G}{\partial\left(\delta f\right)}\frac{\partial f_{0}}{\partial V_{\rm G}},
\label{eq_transcond_diff}
\end{equation}
with the second term in the equation responsible for
the mechanical enhancement.

\begin{figure}[t]
  \begin{center}
	  \begin{tabular}{cc}
      \includegraphics[]{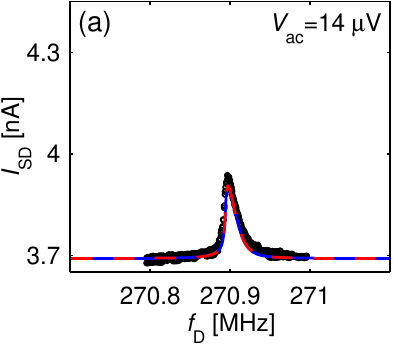}
			\includegraphics[]{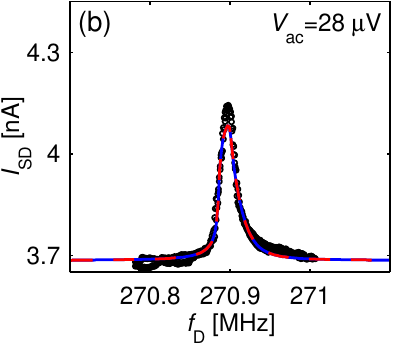}\\
			\includegraphics[]{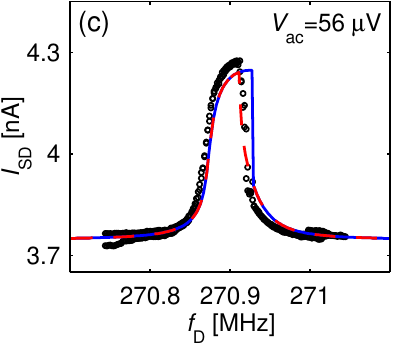}
			\includegraphics[]{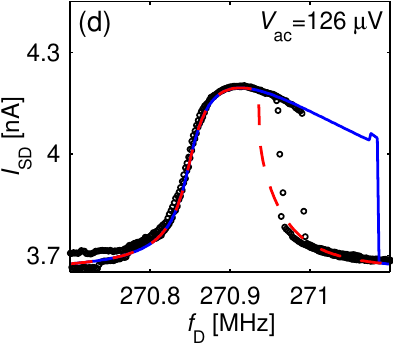}
		\end{tabular}
    \caption{(a)-(d) Source drain dc-current $I_{\rm SD}$ as
        function of drive frequency $f_{\rm D}$. Black circles are
        measured data while lines are theoretical fits. (a) Low
        drive power.  Resonance displaying softening nonlinear
        behavior. (b)-(c) Intermediate drive powers, showing presence
        of a 5$^{\rm th}$ order stiffening nonlinearity. (d) High drive
        power. Mechanically induced gate charge oscillations
        traversing the Coulomb blocade peak results in current
        saturation.}
     \label{fig:nonlinresp}
 \end{center}
\end{figure}

Including higher order terms in the expansion in $x$ of
Equation~\ref{eq:eom1} gives rise to nonlinear terms~\cite{Steele2009b,Lassagne2009,Meerwaldt2012}. From characterizing the resonant response at a point $\Delta V=V_{\rm G0}-V_{\rm deg}=20$~mV from the degeneracy, we find the response of a driven nonlinear resonator
\begin{eqnarray}
&&\ddot{x}+\gamma\dot{x}+\beta x^2\dot{x}+\omega_0^2 x+\alpha x^3+\eta
x^5\nonumber\\
&&=f_{\rm ac}\cos\Omega_{\rm D}t,
\label{eq:effosc}
\end{eqnarray} with parameters
$\gamma/\omega_0=3.45\times10^{-5}$,
$\alpha/\omega_0^2=-1.38\times10^{-4}$~(nm)$^{-2}$,
$\beta/\omega_0=5.64\times10^{-5}$~(nm)$^{-2}$,
$\eta/\omega_0^2=5.18\times10^{-5}$~(nm)$^{-4}$, with
$\omega_0=2\pi\times270.9\times 10^{6}$~s$^{-1}$
and \(\Omega_{\rm D}=2\pi f_{\rm D}\). Note that the Duffing parameter is not exactly canceled out here. In Figure~\ref{fig:nonlinresp}, fits are shown using these parameters together with the measured dc-current for different drive powers as a function of drive frequency. The presence of 5$^{\rm th}$ order nonlinearities is revealed by the change from softening behavior at low drive powers, panel (a), to hardening behavior, panels (c)-(d). Details of the fitting procedure are given in the supplementary material Section S2.

Because the nonlinear parameters have contributions deriving from from single-charge tunneling effects and external electrostatic forces, they are sensitive to the back gate voltage. Hence, although the principle leading up to the simple relation in
Equation~\ref{eq_transcond_diff} remains the same, it becomes more elaborate in the nonlinear regime,
\begin{eqnarray}
&&\frac{dG}{dV_{\rm G}}=\frac{\partial G}{\partial V_{\rm
      G}}+\frac{\partial G}{\partial\omega_0}\frac{\partial
    \omega_{0}}{\partial V_{\rm G}}+\frac{\partial
    G}{\partial\gamma}\frac{\partial \gamma}{\partial V_{\rm G}}
  \nonumber\\ &&+\frac{\partial G}{\partial\alpha}\frac{\partial
    \alpha}{\partial V_{\rm G}}+\frac{\partial
    G}{\partial\beta}\frac{\partial \beta}{\partial V_{\rm
      G}}+\frac{\partial G}{\partial\eta}\frac{\partial \eta}{\partial
    V_{\rm G}}.
\label{eq_transcond_diff2}
\end{eqnarray}

\subsection*{Charge sensitivity.}
To quantify the charge sensitivity, we use procedures practiced for
characterizing SET electrometers. The sensitivity is then calculated from \begin{equation}
\delta Q=\frac{\Delta q_{\mathrm{rms}}}{\sqrt{B}\cdot 10^{\mathrm{SNR}/20}}.
\label{eq_charge_sensitivity_db}
\end{equation}
where \(\Delta q_{\mathrm{rms}}\) is the root mean square gate charge
change caused by a known voltage amplitude reference signal that is
applied to the gate electrode, and \(B\) is the resolution bandwidth
of the measurement device. The signal-to-noise ratio SNR is calculated
as a ratio between the output signal amplitude, measured at the
reference frequency \(f_{\rm R}\), and the noise floor.

We used a reference signal with an amplitude of \(V_{\rm R}=0.025\) mV\subscript{pp} at frequency $f_{\rm R}$ in our measurements. This
signal was fed to the gate electrode of the sample in addition to \(V_{{\rm G}0}\) and $\delta V(t)$. In order to minimize the gate-voltage-source-induced charge noise, the gate was dc-biased with batteries and the sensitivity was measured as a function of the frequency of the rf-drive $f_{\rm D}$. Depending on whether the sample was voltage or current biased, the signal was amplified with a transconductance or voltage amplifier, respectively. The resulting output was recorded with a spectrum analyzer over a frequency range covering \(f_{\rm R}\). It was found that the sensitivity was slightly better when measured in the current bias configuration. The \(V_{\rm{G0}}\) distances to the neighboring Coulomb peaks were 0.632 V and 0.446 V, respectively. For a conservative estimate of the charge sensitivity we thus set the gate period per electron to \(\Delta V_{\rm{G0}}^{e}=0.446\)~V. The reference signal therefore corresponds to rms gate charge change of \( \Delta q_{\rm rms}=e\frac{V_{\rm R}\ \left[\mathrm{V_{\rm rms}}\right]}{\Delta V_{\rm{G0}}^{e}\ \left[\mathrm{V}\right]}=19.8\ \upmu e.\)

\begin{figure}[htb]
\centering
  \begin{tabular}{cc}
    \includegraphics[angle=90]{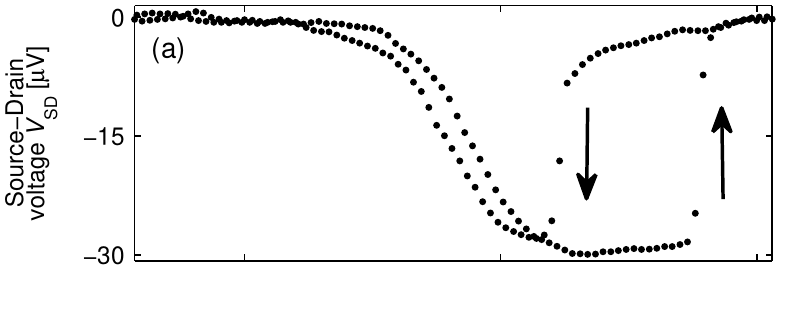}\\[-3ex]
    \includegraphics[]{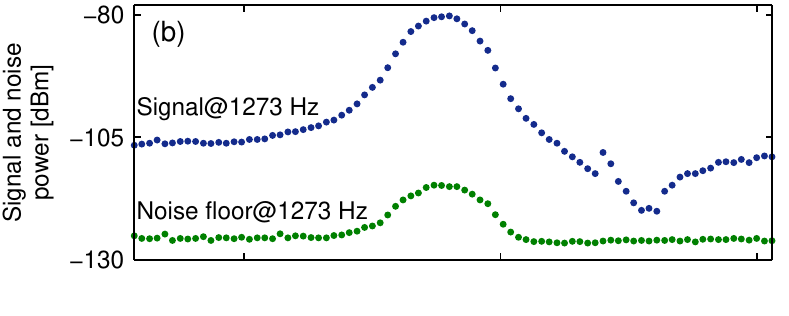}\\[-3ex]
    \includegraphics[]{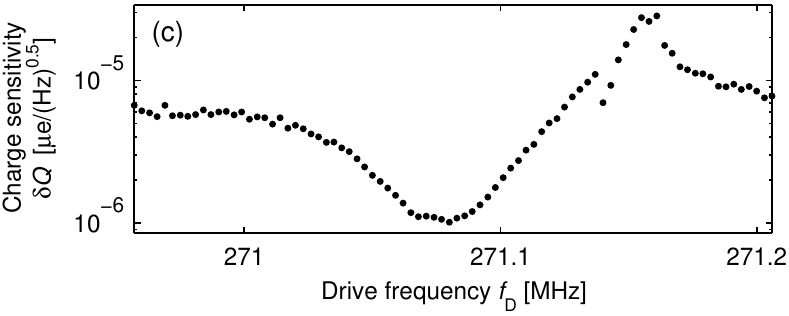}
  \end{tabular}
  \caption{(a) Measured lineshape showing nonlinear hysteretic response. We have subtracted the constant background. Here \(\Delta V_{{\rm G}0}=23\)~mV, \(I_{\rm{SD}}=3\)~nA, and the rms rf-excitation voltage is 71 \(\upmu\)V at the sample. (b) Signal amplitude (blue dots) at the reference frequency $f_{\rm R}=1273$~Hz and the corresponding noise floor (green dots) measured on an upward frequency sweep. There is a small offset between (a) and (b) because the curves were not measured simultaneously. (c) Charge sensitivity $\delta Q$. The best sensitivity is reached by biasing the sample to the left slope of the nonlinear resonance peak.
		}
\label{fig_sensitivity_cbias_1273hz}
\end{figure}

The best mechanical-oscillation-enhanced results were reached by biasing the sample to \(\Delta V_{{\rm G}0}=23\) mV at \(I_{\rm{SD}}=3\)~nA, where the dominating charge transport induced cubic nonlinearity is nearly canceled out. We can see from the measured lineshape in Figure~\ref{fig_sensitivity_cbias_1273hz}(a) that the left side of the peak provides significant \(\left|\partial G/\partial \left(\delta f\right)\right|\) which results in an enhanced overall transducer gain \(\partial V_{\rm SD}/\partial V_{\rm G}\). Charge sensitivity at \(f_{\rm R}=10\)~Hz reaches a minimum of \(\delta Q=4.5\)~\textmu\(e/\sqrt{\mathrm{Hz}}\) at this bias point. In a conventional SET measurement, the best sensitivity is acquired on a bias point where Coulomb effect gives largest \(\partial I_{\rm SD}/\partial V_{\rm G}\). Measuring SNR at the maximum conductance slope of the Coulomb peak in Figure~\ref{fig_f0_vs_Vg}(b) gives a charge sensitivity of \(\delta Q=7.6\)~\textmu\(e/\sqrt{\mathrm{Hz}}\) at \(f_{\rm R}=10\)~Hz, which is worse than in the optimal mechanically enhanced conditions. The transducer gain dependence of the sensitivity implies that the output noise does not solely arise from the noise sources that couple to the island charge in the same way as the reference signal. The experimental results for 10 Hz are displayed in supplementary Figures S6 and S7.

\begin{figure}[ht]
\begin{center}
      \includegraphics[width = 8 cm]{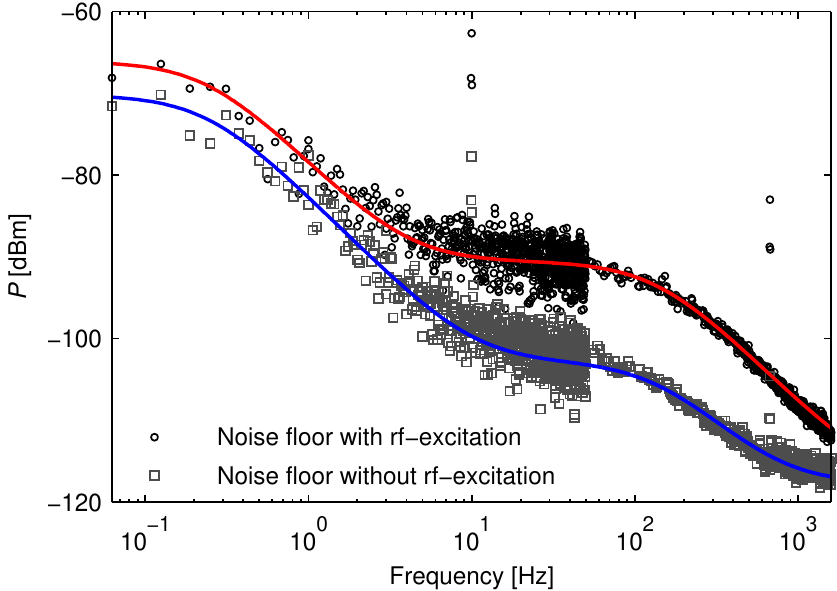}
    \caption{Black circles depict noise floor in current bias configuration at left hand slope of the curve in Figure~\ref{fig_sensitivity_cbias_1273hz}(a). Gray points, squares, show the noise floor at otherwise the same conditions but without rf-excitation. The red and blue lines show qualitative fits acquired from summing two and three Lorentzian power spectra of two-state fluctuators, respectively. Peaks at 10 Hz and 673 Hz are reference signals and correspond to charge sensitivities of 4.5~\textmu\(e/\sqrt{\mathrm{Hz}}\) and 1.1~\textmu\(e/\sqrt{\mathrm{Hz}}\) in driven state, respectively.}
		\label{fig_noise_floor_cbias}
		\label{fig_noise_floor}
\end{center}
\end{figure}

To enhance the sensitivity further, the easiest is to exploit \(1/f^{\alpha}\) noise dependence and increase the operation frequency $f_{\rm R}$. Maximum operation frequency is limited by the RC filter capacitors and the sample resistance which set the cut-off frequency of the circuitry to around two kHz. The noise floor behavior is shown in Figure~\ref{fig_noise_floor} up to \(1.6\) kHz when measured at the left slope of the resonance curve in Figure~\ref{fig_sensitivity_cbias_1273hz}(a). In the driven state the noise characteristics can be qualitatively reproduced by summing two Lorentzian power spectra of two-level fluctuators (TLF).~\cite{Tarkiainen2005} Between \(\sim200\) Hz and \(1.6\) kHz the output noise is dominated by a single TLF and the noise rolls off as \(1/f^2\). Due to the limited bandwidth of the measurement circuitry, the signal-to-noise ratio improves only up to \(\sim1.3\) kHz.

Figure~\ref{fig_sensitivity_cbias_1273hz}(b) shows results of a current biased charge sensitivity measurement at \(f_{\rm R}=1273\)~Hz when \(f_{\rm D}\) is swept upward. Blue dots are the signal amplitude at $f_{\rm R}$ while the noise floor is represented by green dots. The behavior of the signal suggests that the additional terms stemming from the nonlinearity included in Equation~\ref{eq_transcond_diff2} are not dominant. In addition, a discontinuity in the signal can be seen around 271.16 MHz where the instability point is reached. To obtain the charge sensitivity in Figure~\ref{fig_sensitivity_cbias_1273hz}(c), the SNR is calculated for each data pair of the signal (blue dot) and noise power (green dot). From Figure~\ref{fig_sensitivity_cbias_1273hz}(c) it is seen how, despite the increase in noise, the driving of the mechanical motion enhances the charge sensitivity from \(\delta Q=6\)~\textmu\(e/\sqrt{\mathrm{Hz}}\) approximately by a factor of six down to the best sensitivity of \(\delta Q=0.97\)~\textmu\(e/\sqrt{\mathrm{Hz}}\).

From the green dots in Figure~\ref{fig_noise_floor}(b), an increase of the noise floor is seen in the vicinity of the maximum signal. While we abstain from an in-depth analysis of this noise, the fact that we find an optimum sensitivity when the resonator is biased such that one of the nonlinearities vanish, partly confirms the important role played by the nonlinearities in the system. A detailed analysis of the relation between the resonance and the observed noise is here made complicated by the numerous nonlinearities present. In our case, the effective oscillator equation contains both conservative as well as dissipative nonlinearities which should be supplemented by corresponding additive and multiplicative noise. We also note that the \(V_{\rm G}\)-dependence of these coefficients makes them couple to the charge noise. Further, correlations between single electron tunnel events and resonator motion, known to affect the \textit{Q}-factor of these devices, can also influence the noise.

\subsection*{Summary and conclusion.} In conclusion, we have studied how mechanical oscillation can be used for improving the charge sensitivity of a carbon nanotube QD. Driving the mechanical resonator far into the nonlinear regime causes a relatively large change in the low frequency conductance of the device. Because the charge-tunneling-induced spring constant softening makes \(df_{0}/dV_{\rm G}\) enhanced, the conductance slope change causes transducer gain to increase as implied by nonzero \(\frac{\partial G}{\partial\left(\delta f\right)}\frac{\partial f_{0}}{\partial V_{\rm G}}\).

We find best charge sensitivities at a bias point where 5\(^{\rm th}\) order nonlinearity dominates the high amplitude mechanical motion. A sensitivity of \(\delta Q=4.5\)~\textmu\(e/\sqrt{\mathrm{Hz}}\) at 10~Hz is slightly better than previously measured with low frequency SETs.~\cite{Krupenin2000} Operation at 1273~Hz yields optimal charge sensitivity \(\delta Q=0.97\)~\textmu\(e/\sqrt{\mathrm{Hz}}\). The charge sensitivity enhancement due to driven mechanical motion amounted to a factor of six.

For comparison, the previous record charge sensitivity using a SWNT rf-SET
was 2.3~\textmu\(e/\sqrt{\mathrm{Hz}}\) and a metallic rf-SET \(\delta Q=0.9\)~\textmu\(e/\sqrt{\mathrm{Hz}}\).~\cite{Andresen2008,Brenning2006} Latter device was operated in the superconducting state at 40 mK. It should be noted that Brenning et al.~\cite{Brenning2006} calculated the sensitivity of their record holding device assuming the possibility of using information from both sidebands. Thus, the charge detector presented in this letter provides a charge sensitivity that is approximately a factor $1/\sqrt{2}$ better than actually measured previously.

\subsection*{Associated content}
Supplementary information. Detailed description of the characterization of electrical and mechanical properties, the fitting procedure used for the mechanical response curves, and additional data on the mechanical \(5^{\rm th}\) order nonlinearity are provided. This material is available free of charge via the Internet at http://pubs.acs.org.


\begin{acknowledgement}
We acknowledge fruitful discussions with A. Bachtold. Our work was supported by the Academy of Finland (contract no. 250280, LTQ CoE), Swedish research council (VR), and the European Union Seventh Framework Programme under grant agreement no 604391 Graphene Flagship. We benefitted from the use of the Aalto University infrastructures Low Temperature Laboratory and Nanomicroscopy Center (Aalto-NMC).
\end{acknowledgement}

\providecommand{\latin}[1]{#1}
\providecommand*\mcitethebibliography{\thebibliography}
\csname @ifundefined\endcsname{endmcitethebibliography}
  {\let\endmcitethebibliography\endthebibliography}{}

\onecolumn
\input{charge_sensitivity_nanolett_supp.tex}

\end{document}

%% file: charge_sensitivity_nanolett_supp.tex
\setcounter{figure}{0}
\renewcommand{\thefigure}{S\arabic{figure}}
\setcounter{equation}{0}
\renewcommand{\theequation}{S\arabic{equation}}
\setcounter{table}{0}
\renewcommand{\thetable}{S\arabic{table}}
\section*{S1. Charge dynamics and electrical characterization}
The single electron tunneling of units of charge $\left(e=|e|>0\right)$ to and from the CNT-dot is modeled using the orthodox theory for Coulomb Blockade. In this regime, the charge transfer rates are given by the
expressions
\begin{equation}
\Gamma_{\rm L,R}^{\pm e}=\frac{\Gamma_{\rm L,R}^{(0)}}{\Delta_{\rm D}}\frac{\Delta U_{\rm L,R}^{\pm e}}{1-\exp{\left(-\beta\Delta U_{\rm L,R}^{\pm e}\right)}},
\label{eq:ortho}
\end{equation}
where $\Delta_{\rm D}$ is the level width of the state on the dot.  In Coulomb blockade regime, in the vicinity of charge degeneracy points, the change in free energy $\Delta U$ is most conveniently expressed in the two-state model where transitions between charge states $N$ and $N+1$ dominate. For a given bias configuration on the left (L), right (R) and gate (G) electrodes one finds using Eq.~\ref{eq:ortho}
\begin{eqnarray}
&&\Gamma_{\rm L}^+(N\rightarrow N+1)=\frac{\Gamma_L^{(0)}}{\Delta_{\rm D}}f(\Delta-eV_{\rm S}),\\
&& \Gamma_{\rm L}^-(N+1 \rightarrow N)=\frac{\Gamma_L^{(0)}}{\Delta_{\rm D}}f(-\Delta+eV_{\rm S}),\\
&& \Gamma_{\rm R}^-(N+1 \rightarrow N)=\frac{\Gamma_R^{(0)}}{\Delta_{\rm D}}f(-\Delta+eV_{\rm D}),\\
&& \Gamma_{\rm R}^+(N \rightarrow N+1)=\frac{\Gamma_R^{(0)}}{\Delta_{\rm D}}f(\Delta-eV_{\rm D}).
\end{eqnarray}
Here $f(E)=E/\left[\exp\left(E/k_{\rm B}T^{*}\right)-1\right]$,
$\Delta=\frac{e}{C_{\Sigma}}\left[Ne+\frac{e}{2}+{Q^*}+{V_{\rm G}C_{\rm G}(x)}\right]$, and \(V_{\rm S}\) and \(V_{\rm D}\) are source and drain voltages, respectively. $C_{\Sigma}$ is the total capacitance that comprises of $C_{\rm S}$, $C_{\rm D}$, $C_{\rm G}$, and $C_0$ which are source, drain, gate, and ground capacitances, respectively. Here $Q^*=V_{\rm S}C_{\rm S}+V_{\rm D}C_{\rm D}$ is assumed to be independent of resonator deflection $x$. In terms of these parameters, the two state approximation implies $-(N+1)e<V_{\rm G}C_{\rm G}+Q^* < -Ne.$

As the typical currents in the experiments are in the nA-regime, the
average charge transfer rate is $\Gamma\sim 10^{-9}\ \rm{A}/e\approx 6$ GHz which
is much larger than the frequency of mechanical vibration in the device. In the limit
$\Gamma>> \omega_0$, one can make an adiabatic approximation to find
the charge on the tube and the current. In this approximation, one has
\begin{eqnarray}
&&\left<Q(t)\right>\approx Ne+\frac{\Gamma_{\rm R}^++\Gamma_{\rm L}^+}{\Gamma_\Sigma}e,\quad I\approx \frac{e}{\Gamma_\Sigma}\left(\Gamma_{\rm L}^+\Gamma_{\rm R}^--\Gamma_{\rm R}^+\Gamma_{\rm L}^-\right),
\end{eqnarray}
where $\Gamma_\Sigma=\Gamma_{\rm L}^++\Gamma_{\rm R}^++\Gamma_{\rm L}^-+\Gamma_{\rm R}^-$.

\subsection*{S1.1 Static operation and device characterization}
\begin{figure}[htb]
\centering
\includegraphics[width=1\linewidth]{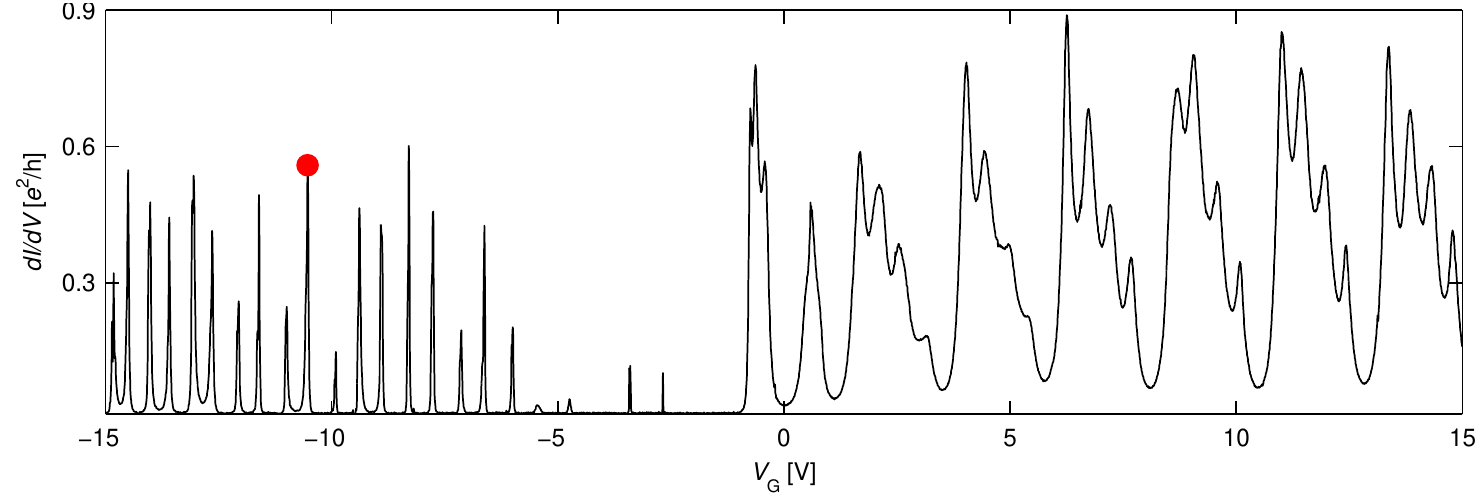}
\caption{Differential conductance of the carbon nanotube device is plotted as function of $V_{\rm G}$ at $V_{\rm SD}=1$~mV. Contacts are more transparent at positive gate voltages. The four fold degeneracy is also clearly visible at electron conduction regime. The red circle indicates the degeneracy point where charge sensitivity measurements have been carried out.
\label{fig:CB_peaks}}
\end{figure}

\begin{figure}[htb]
\centering
\includegraphics[width=1\linewidth]{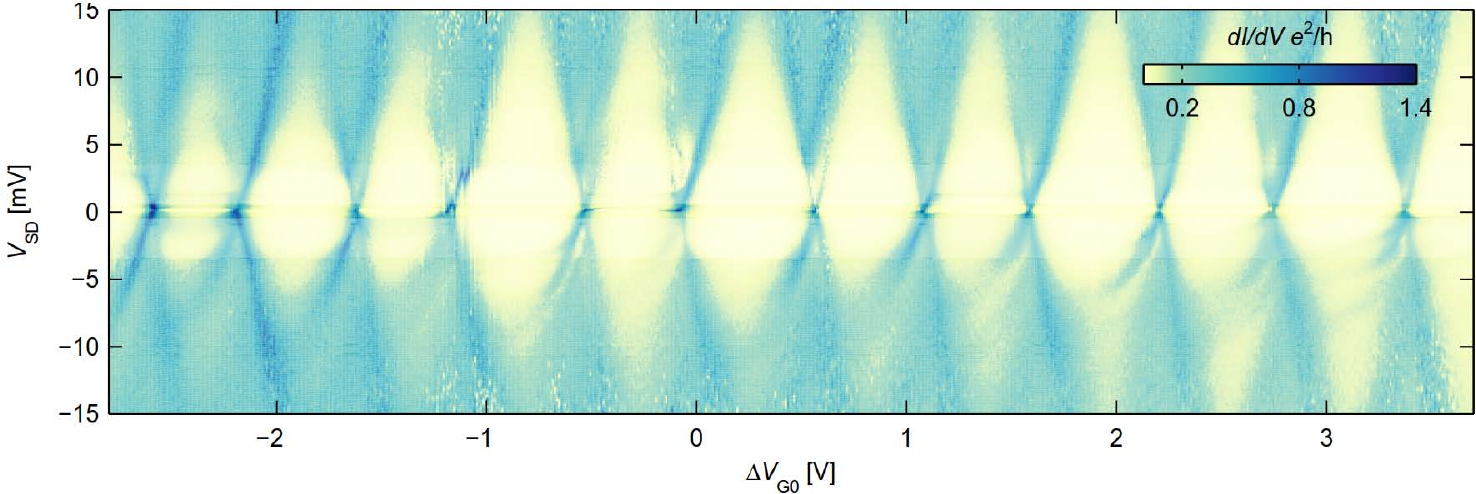}
\caption{Differential conductance as function of $V_{\rm G}$ and $V_{\rm SD}$ is plotted around the charge degeneracy highlighted in Figure~\ref{fig:CB_peaks}, \(\Delta V_{\rm{G0}}=V_{\rm{G0}}-V_{\rm{deg}}\). All of the Coulomb diamonds display an early onset of charge transport, which is seen as nonzero conductance appearing before the diamonds are completed at high \(\left|V_{\rm{SD}}\right|\) values. This is caused by contunneling processes arising from enhanced tunnel coupling.
	\label{fig:CB_diamonds}}
\end{figure}

\begin{figure}[htb]
\centering
  \begin{tabular}{cc}
    \includegraphics[]{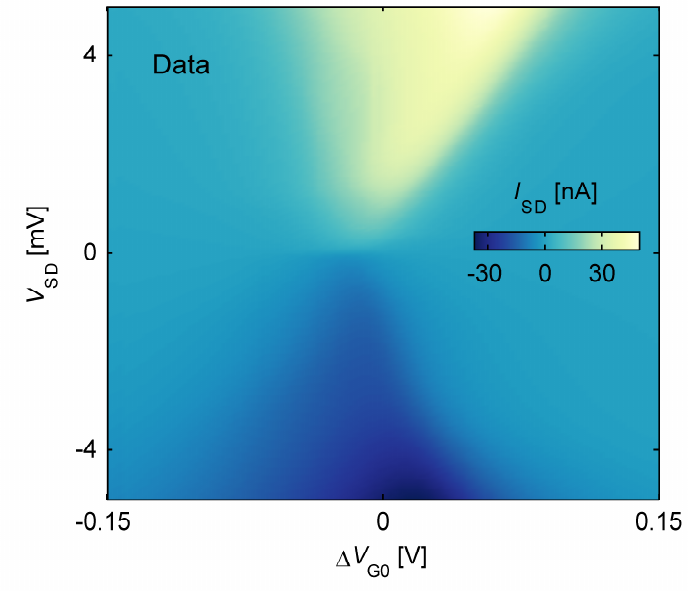} &
    \includegraphics[]{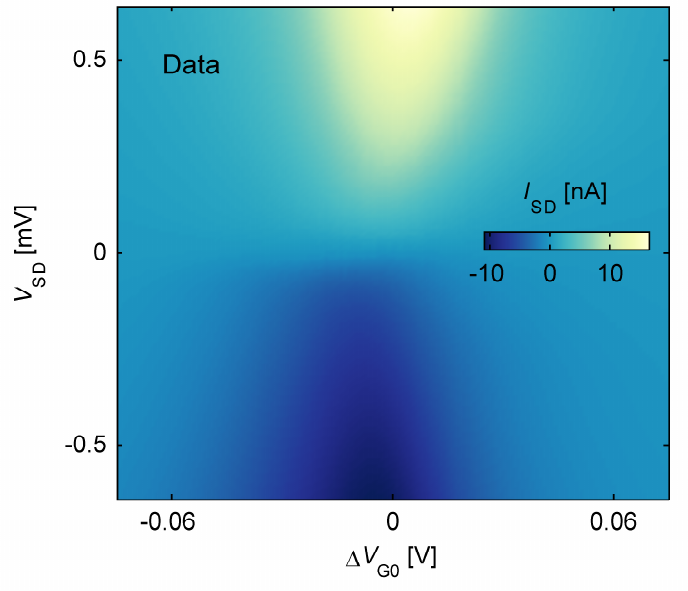}   \\
    \includegraphics[]{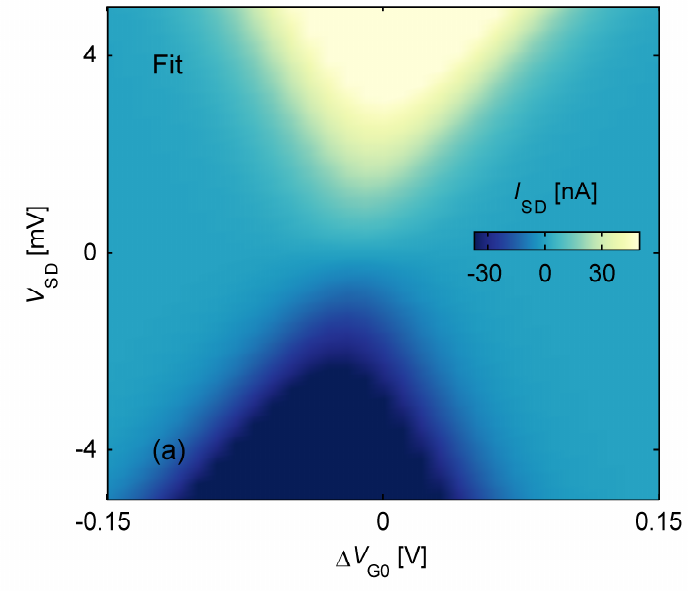} &
    \includegraphics[]{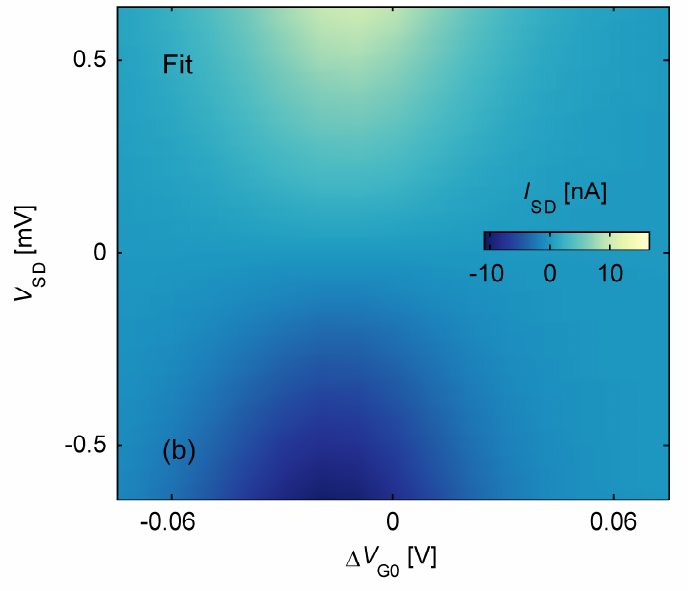}
  \end{tabular}
  \caption{Measured (top) and calculated (bottom) currents in the vicinity of the
charge degeneracy point at $V_{\rm G}=V_{\rm deg}\approx-11.5$~V. (a) and (b)
show two different ranges of the source-drain bias voltage. The gate voltage scale has been offset from the charge degeneracy, \(\Delta V_{\rm{G0}}=V_{\rm{G0}}-V_{\rm{deg}}\).}
\label{fig:cap_fit}
\end{figure}

Electrical characterization of the device shows clear Coulomb
blockade behavior, see Figures~\ref{fig:CB_peaks} and \ref{fig:CB_diamonds}. The measurements of mechanical resonance and charge sensitivity were carried out near the degeneracy point marked by the red circle at $V_{\rm G}=V_{\rm deg}\approx-11.5$~V in Figure~\ref{fig:CB_peaks}. To obtain electrical device parameters, more detailed measurements were done near the degeneracy point. Measured and fitted data are shown together in Figure~\ref{fig:cap_fit} and the modeling parameters are summarized in Table~\ref{tab:cap}.

%The fitting parameters
%$T=4.5$~K, $C_0=1$~aF, $C_{\rm S}=3.4$~aF, $C_{\rm D}=7.2$~aF, $C_{\rm G}=0.34$~aF, and $\Gamma_{\rm L,R}/\Delta_{\rm D}=0.25\times 10^{34}$~s$^{-1}$J$^{-1}$.

\begin{table}[h]
\caption{Fitted circuit parameters around the degeneracy point at $V_{\rm G}=V_{\rm deg}$. The effective electronic temperature is high because the lifetime broadening in strong tunneling regime is neglected in the model.\label{tab:cap}}
\begin{tabular}{l|r}
\hline\hline
Quantity  & Value \\
\hline
Temperature, $T^{*}$ & 4.5 K\\
Source capacitance, $C_{\rm S}$ & 3.4 aF\\
Drain capacitance, $C_{\rm D}$ & 7.2 aF\\
Gate capacitance, $C_{\rm G}$ & 0.34 aF\\
Stray capacitance, $C_0$  & 1.0 aF\\
Rate constant, $\Gamma_{\rm L,R}/\Delta_{\rm D}$ & $0.25\times 10^{34}$~s$^{-1}$J$^{-1}$\\
\hline
\end{tabular}
\end{table}

\clearpage

\section*{S2. Mechanical subsystem}
First we derive the force experienced by the carbon nanotube quantum dot. Denote the electrical potential on the carbon nanotube dot by $\phi$. If the gate, source, and drain are connected to reservoirs held at fixed potentials the potential on the dot can be expressed in terms of the net average charge $Q(t)$ on the dot as
$$\phi=\frac{Q(t)+Q^*+V_{\rm G}C_{\rm G}}{C_\Sigma}.$$
The total free electrostatic energy of the system is for given $Q(t)$ \footnote{For a system of charged conductors the electrostatic energy is $\frac{1}{2}\sum_n Q_nV_n$. If a subset of the conductors $n\in A$ are held at fixed potentials, the redistribution of charge from the associated reservoirs must be taken into account. The force acting on a conductor $j$ is thus given by $F_j=-\partial {\cal F}/\partial x_j$, where ${\cal F}$ is the free energy ${\cal F}=\frac{1}{2}\sum_{n} Q_nV_n-\sum_{n\in A} Q_nV_n$. (See for instance chap.2 in {\it Static and Dynamic Electricity} by W. R. Smythe, 2$^{\rm nd}$ ed., McGraw-Hill, New-york, 1950.)}
$${\cal F}=\frac{1}{2}[Q(t)\phi-V_{\rm S}(V_{\rm S}-\phi)C_{\rm S}-V_{\rm D}(V_{\rm D}-\phi)C_{\rm D}-V_{\rm G}(V_{\rm G}-\phi)C_{\rm G}].$$
The force is determined from the relation $F=-\partial{\cal F}/\partial x$. If the only the gate capacitance has an $x$ dependence one obtains
$$F=\frac{1}{2}[V_{\rm G}(V_{\rm G}-\phi)C_{\rm G}']-\frac{1}{2}[Q(t)+Q^*+V_{\rm G}C_{\rm G}]\frac{\partial \phi}{\partial x},$$
where $C_{\rm G}'=d C_{\rm G}/d x$. Noting that ${\partial \phi}/{\partial x}={C_{\rm G}'}{C_\Sigma^{-1}}(V_{\rm G}-\phi),$ the force can be written as
$$F=\frac{1}{2}[V_{\rm G}(V_{\rm G}-\phi)C_{\rm G}']-\frac{1}{2}\phi{C_{\rm G}'}(V_{\rm G}-\phi)=\frac{1}{2}C_{\rm G}'(V_{\rm G}-\phi)^2.$$
In terms of the charge on the dot $Q(t)$, this becomes
$$F=\frac{1}{2}C_G'V_{\rm G}^2\left(1-\frac{Q(t)+Q^*+V_{\rm G}C_{\rm G}}{V_{\rm G}C_\Sigma}\right)^2.$$

Considering only very small vibrations around the resonator equilibrium position (amplitudes $\lesssim$ 1 nm). The equation of motion for the fundamental flexural mode coordinate $x$ of the resonator, can be written as
\begin{equation}
\ddot{x}+\gamma_0\dot{x}+\Omega_0^2 x=\frac{1}{2m}C_{\rm G}'(x)V_{\rm G}^2\left(1-2\frac{\left<Q(t)\right>+Q^*+V_{\rm G}C_{\rm G}(x)}{V_{\rm G}C_\Sigma}\right),
\label{eq:eom}
\end{equation}
where \(\Omega_0\) is resonance frequency at zero gate bias, \(\Omega_0/\gamma_0\) the quality factor, and \(m\) effective mass of the resonator. We have here invoked the adiabatic approximation replacing $Q(t)\rightarrow \left<Q(t)\right>$ and neglected terms of second order in the dot potential since we have \(\left|\left[\left<Q(t)\right>+\right.Q^*+\left.V_{\rm G}C_{\rm G}(x)\right]/{V_{\rm G}C_\Sigma}\right|\ll1\) at the chosen degeneracy point.

\subsection*{S2.1 Frequency tuning of mechanical resonance}
\begin{figure}[t]
\centering
\includegraphics[]{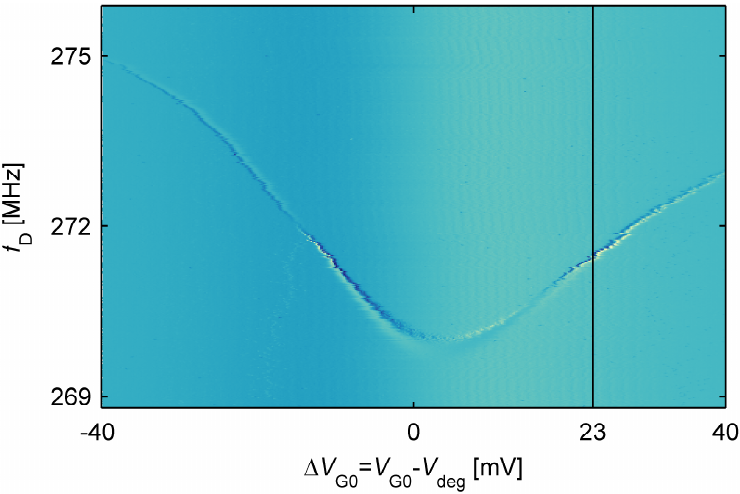}
\includegraphics[]{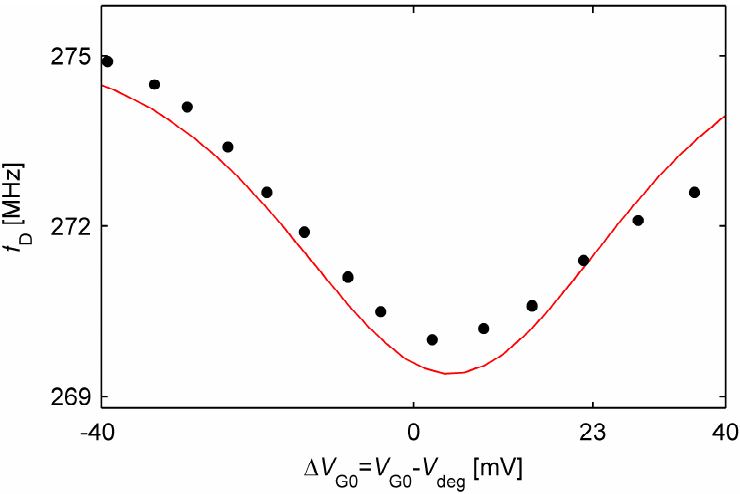}
\caption{Measured and fitted dynamic frequency tuning near the degeneracy point at $V_{\rm G0}=V_{\rm deg}$. (a) False color image of transconductance as function of drive frequency and back gate voltage (offset from the degeneracy point) shows the mechanical resonance as a faint line on the background. Vertical line shows the gate bias point used for charge sensitivity experiments. (b) Theoretical curve for dynamic softening given by Equation 3 of the main text  is plotted with the red line.
	}
	\label{fig:tune_dyn}
\end{figure}

To characterize the mechanical resonance, the large scale frequency
tuning with back gate voltage was considered. Measured data points are
shown as black markers in Figure~2(a) of the main text together with a
least square fit of a parabolic dependence of frequency on back gate
voltage. To relate the tuning to model parameters, we expand the
right hand side of Eq.~\ref{eq:eom} to lowest order in the
displacement $x$,
\begin{equation}
\ddot{x}+\gamma_0\dot{x}+\Omega_0^2 x\approx\frac{1}{2m}(C_{\rm G}'+xC_{\rm G}'')V_{\rm G}^2-x\frac{V_{\rm G}^2C_{\rm G}'^2}{mC_{\rm G}C_\Sigma}\frac{\partial \left<Q(t)\right>}{\partial V_{\rm G}}.
\label{eq:tune}
\end{equation}
The first term on the right hand side of Eq.~\ref{eq:tune} provides the large scale frequency tuning with back gate voltage while the second term is responsible for the dynamic softening in the vicinity of the degeneracy points. From the fit of the global tuning in Figure 2(a) we find ${a=C_{\rm G}''}/{2m}\approx 2.83\times 10^{15}$~(Vs)$^{-2}$. Fitting to the tunneling induced frequency tuning measured at \(V_{G0}=V_{\rm deg}\approx-11.5\) V Coulomb peak yields $b=C_{\rm G}'^2/m\approx 1.4\times 10^{-3}$~F$^2$/m$^{2}$kg. Data and the fit are shown in Figure~\ref{fig:tune_dyn}.

On dimensional grounds, it follows that $m \propto b/(ah)^2$, where \(h=220\) nm is the height of the vacuum gap between the suspended tube and the back gate dielectric. The constant of proportionality depends on the capacitance model used for the estimate. We find an effective mass ranging from \(9.8\times 10^{-22}\) kg (thin wire over plate) to \(3.6\times10^{-21}\) kg (parallel plate model). For concreteness, we have in the numerical fitting to the resonance curves used a model of a wire over a plane
\[
C_{\rm G}=\frac{2\pi\epsilon_0\epsilon_r L}{{\rm acosh}(h/r)}
\]
with an adjusted effective dielectric constant \(\epsilon_{\rm r}=0.113\), suspended length \(L=280\) nm, and air gap height \(h=220\) nm, which reproduce the measured gate capacitance value \(C_{\rm G}=0.34\) aF with a diameter of \(2r=d=5\) nm. This corresponds to an effective mass of \(m=1.7\times 10^{-21}\) kg.

\clearpage
%\noindent {\bf Reality check}: In parallel plate capacitor model $C_{\rm G}'\approx -C_{\rm G}/d$ and $C_{\rm G}''\approx 2C_{\rm G}/d^2$. This would give $C_{\rm G}\approx 1.4\times 10^{-3}/2.8\times 10^{-15}\approx 0.5$~aF, which is close to the used fitting value of .34 aF.
%===============================================================================================================================================
\subsection*{S2.2 Resonant response}
To probe the resonant response, an ac-signal was added to the gate with
varying ac-power. Sweeping the drive frequency up and down around the
resonance, the resonant response was measured by monitoring the
dc-current, time averaged current $I_{\rm SD}=\left<I(t)\right>$,
through the device. The time dependence of the current comes mainly
from the dependence of the island charge on the gate charge $V_{\rm G}(t)C_{\rm G}(x(t))$. In order to characterize the mechanical response, the stationary current measured as a function of gate voltage was first fitted to a Lorentzian shape in the neighborhood of the degeneracy point. $$I(V_{\rm G})\approx \frac{I_0}{1+(V_G-V_{\rm G0})^2/\Delta V^2}.$$
with fitting parameters $V_{\rm G0}=V_{\rm deg}=-11.5381$~V, $I_0=7.4$~nA, and
$\Delta V=20.1$~mV.

\begin{figure}[ht]
\centering
	  \begin{tabular}{cc}
      \includegraphics[]{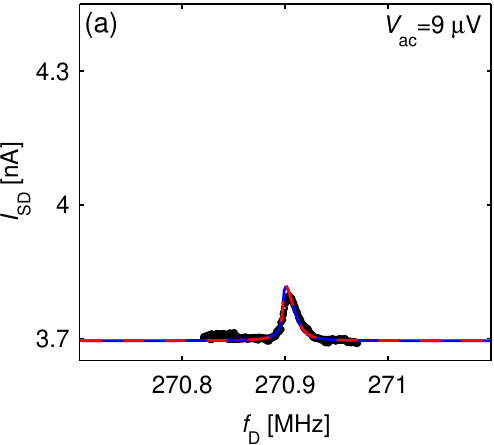}
			\includegraphics[]{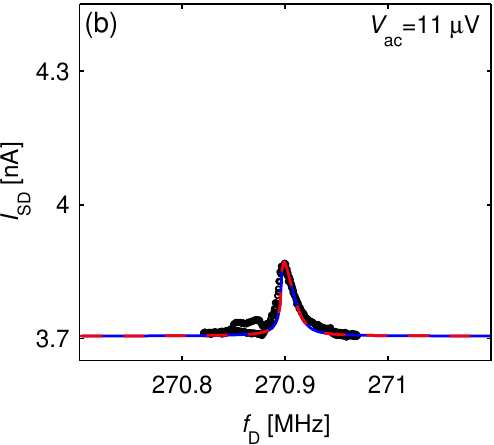}
			\includegraphics[]{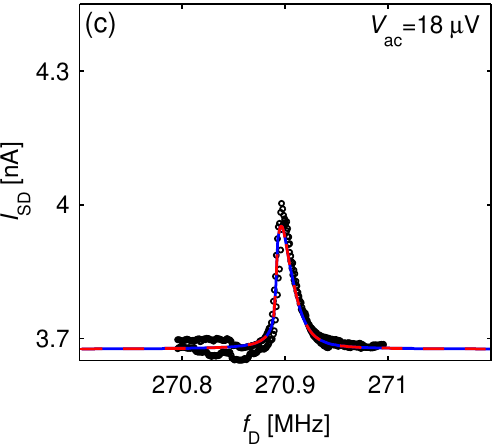}\\
      \includegraphics[]{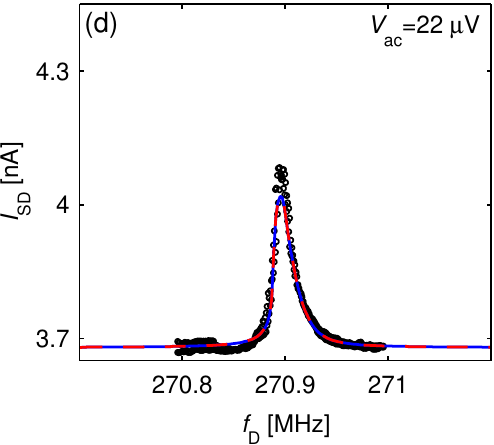}
			\includegraphics[]{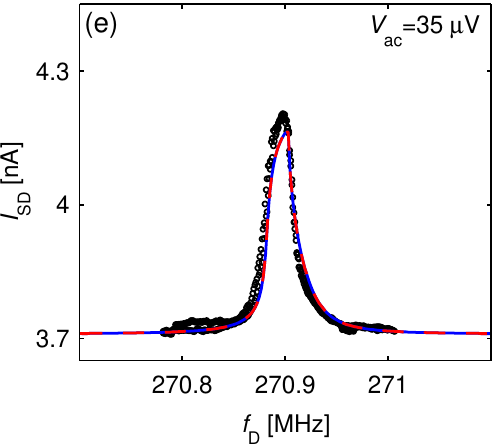}
			\includegraphics[]{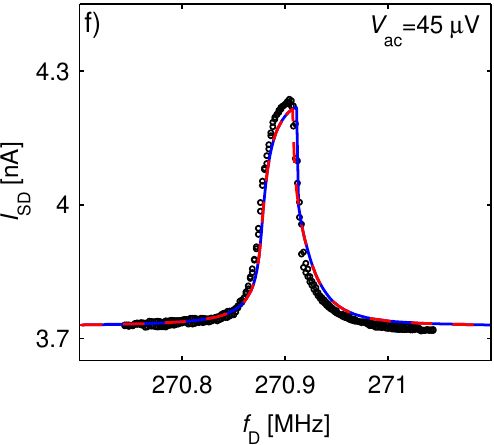}\\
      \includegraphics[]{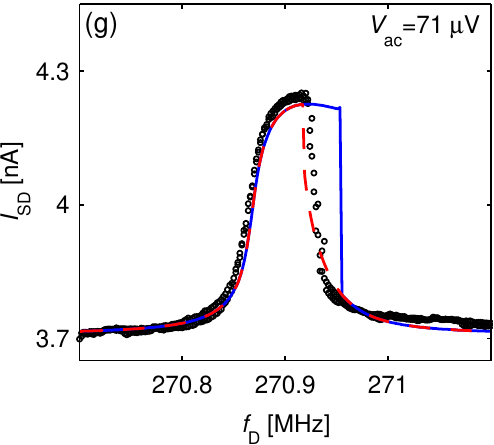}
			\includegraphics[]{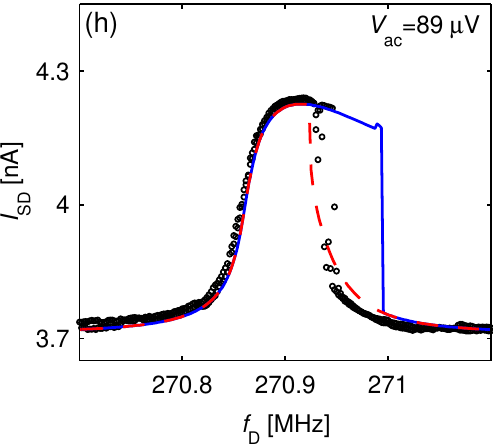}
			\includegraphics[]{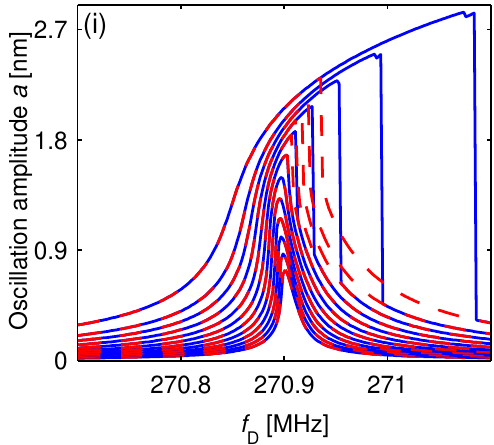}
			%\\[-0.5ex]
		\end{tabular}
		\caption{Measured dc-currents $I_{\rm SD}$ as function of drive frequency $f_{\rm D}$ at $\Delta V_{\rm G0}=20$~mV. Panels (a)-(h) correspond to increasing drive powers $V_{\rm ac}$.  Solid blue lines are the calculated currents based on the model for upwards frequency sweeps while dashed red lines correspond to calculated downward frequency sweeps. Panels (a)-(c) show clear softening behavior indicative of a negative Duffing constant. At higher drives the response is dominated by a stiffening fifth order nonlinearity; panels (f)-(h). Current saturation occurs at large amplitudes as the dot chemical potential traverses the degeneracy point during the large amplitude vibrations; panels (f)-(h). (i) Oscillation amplitude $a$ in nm as function of drive frequency corresponding to measured curves shown in panels (a)-(h) and curves shown in Figure 3 of the main text. Blue solid lines correspond to upward frequency sweeps while red dashed lines to downward sweeps. Note that although the current saturates, the amplitudes are still monotonically increasing with higher drive.
	\label{fig:nlinresp}}
\end{figure}

With $V_{\rm G}=V_{\rm deg}\left[1+\epsilon_{\rm offs}+\epsilon_{\rm ac}\cos\left(\Omega_{\rm D} t\right)\right]$, \(\Omega_{\rm D}=2\pi f_{\rm D}\), and a nearly harmonic stationary response $x(t)\approx a\cos(\Omega_{\rm D}t+\phi)$ this results in a dc-current
\begin{equation}
I_{\rm
  dc}(a,\phi)=\frac{I_0}{2\pi}\int_0^{2\pi}d\nu\,\left[1+\frac{V_{\rm
      deg}^2}{\Delta V^2}(\epsilon_{\rm offs}+\epsilon_{\rm
    ac}\cos\nu+\frac{C_G(a\cos[\nu+\phi])}{C_G(0)}-1)^2\right]^{-1}.\label{eq:avcurr}
\end{equation}
Here \(\epsilon_{\rm offs}=\Delta V_{\rm G0}/V_{\rm deg}\) and \(\epsilon_{\rm ac}=V_{\rm ac}/V_{\rm deg}\). The integration is made over one oscillation period. As can be seen from Figure~\ref{fig:nlinresp}, the response is strongly nonlinear. To fit the resonance with drives $V_{\rm ac}>25~\upmu{\rm V}$, a model with cubic \(\alpha x^3\) and pentic \(\eta x^5\) conservative nonlinearities as well as a third order nonlinear damping term \(\beta x^2\dot{x}\) is necessary. This corresponds to a model resonator equation
$$\ddot{x}+\gamma\dot{x}+\beta x^2\dot{x}+\omega_0^2 x+\alpha x^3+\eta
x^5=f_{\rm ac}\cos\Omega_{\rm D}t,$$
where \(\gamma\) is the regular linear damping coefficient. In the rotating wave approximation, the corresponding equations for the stationary state amplitude $a$ and phase $\phi$ read
\begin{eqnarray}
&&a^2\left(\left[\omega_0^2-\Omega_{\rm
      D}^2+\frac{3\alpha}{4}a^2+\frac{5\eta}{8}a^4\right]^2+\Omega_{\rm
    D}^2\left[\gamma+\frac{\beta}{4}a^2\right]^2\right)=f_{\rm
    ac}^2,\label{eq:aeq}\\ &&\tan \phi=\frac{\Omega_{\rm
      D}[\gamma+\frac{\beta}{4}a^2]}{\Omega_{\rm
      D}^2-\omega_0^2-\frac{3\alpha}{4}a^2-\frac{5\eta}{8}a^4}\label{eq:phieq}.
\end{eqnarray}

In Figures~\ref{fig:nlinresp}(a)-(h), the solid and dash dotted lines result from fitting the model parameters $\gamma$, $\beta$, $\alpha$, and $\eta$ to a set of measured data at an offset bias of $\Delta V_{\rm G0}=20$~mV. A total of 24 different powers corresponding to drive voltages in the interval $9~\upmu{V}<V_{\rm ac}<126~\upmu{V}$ were fitted with the same parameter set. The fits correspond to the parameters
$\gamma=3.45\times10^{-5}\omega_0$,
$\alpha=-1.38\times10^{-4}\omega_0^2$~(nm)$^{-2}$,
$\beta=5.64\times10^{-5}\omega_0$~(nm)$^{-2}$,
$\eta=5.18\times10^{-5}\omega_0^2$~(nm)$^{-4}$, with $\omega_0=2\pi\times270.9\times 10^{6}$~Hz.

The fitting was done by solving numerically the
Equations~\ref{eq:aeq} and \ref{eq:phieq} and then calculating the
integral in Equation \ref{eq:avcurr}. Considering first the low drive data, panels (a)-(c), the negative Duffing constant $\alpha$ was determined to reproduce the backbone curve. Then, the fifth order nonlinearity $\eta$ was added to reproduce the backbone in panels (f)-(h). The linear damping coefficient $\gamma$ was chosen to reproduce the width of the resonances, while the nonlinear damping coefficient $\beta$ was fitted to obtain agreement with the observed width of the hysteresis.

\clearpage

\begin{figure}[ht]
\centering
	  \begin{tabular}{cc}
      \includegraphics[]{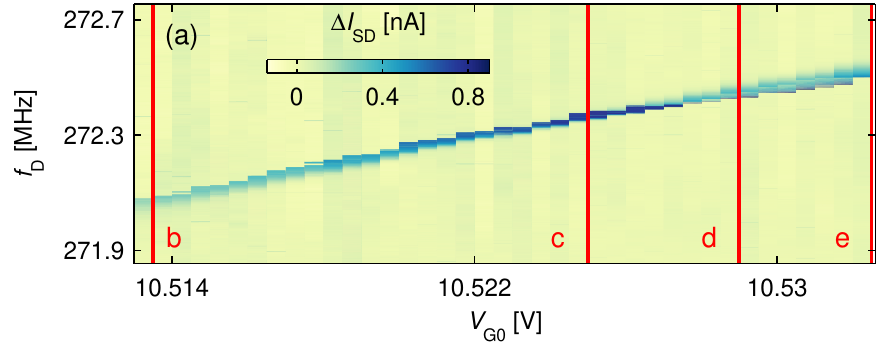}\\
			\includegraphics[]{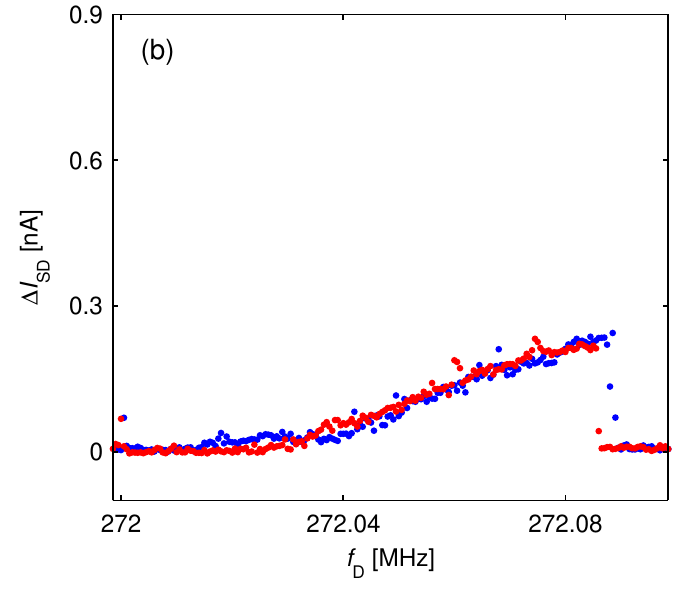}
			\includegraphics[]{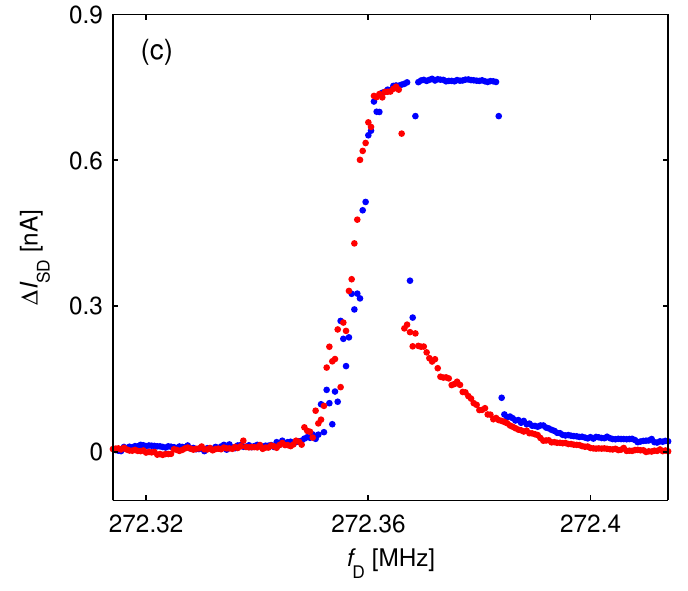}\\
      \includegraphics[]{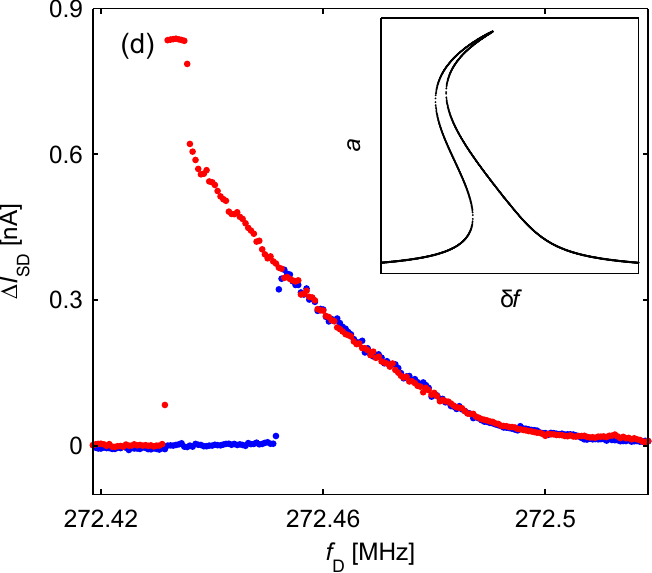}
      \includegraphics[]{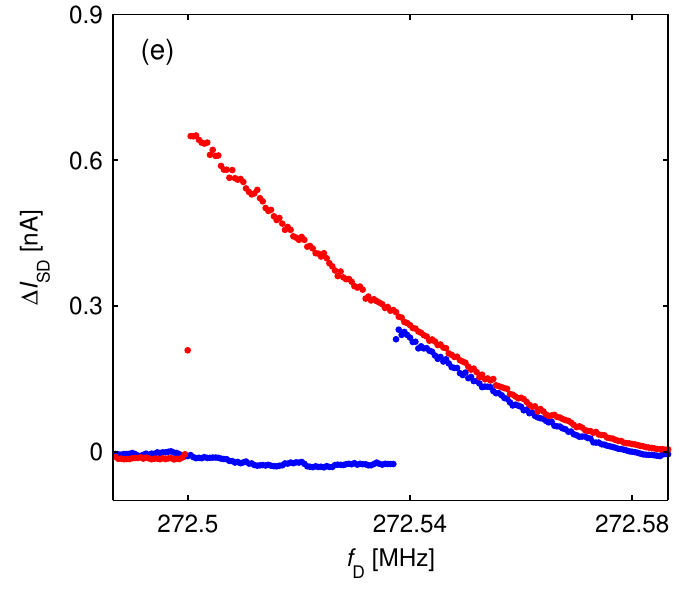}
			%\\[-0.5ex]
		\end{tabular}
		\caption{These data were measured during a separate cool down at a charge degeneracy point located at \(V_{\rm G0}\approx 10.5\) V. The measurement was made with \(V_{\rm SD}=0.25\) mV and \(V_{\rm ac}=40\) \textmu V. The change of the resonance frequency is plotted in (a) against \(V_{\rm G0}\) around the right \(\alpha=0\) point. Vertical lines highlight the positions of the cross sectional resonance response curves displayed in (b)-(e). Upward sweeps are denoted in blue and downward sweeps in red. At the furthest ends of the presented \(V_{\rm G0}\) range, Duffing parameter takes a positive (b) and negative (e) value due to a sign change of the single-electron-tunneling-induced Duffing constant. Close to the \(\alpha=0\) point, a small \(3^{\rm rd}\) order term allows the \(5^{\rm th}\) order term to dominate the large amplitude behavior. The resulting response in (c) is similar to Figure~\ref{fig:nlinresp}. When the Duffing constant decreases further, the oscillation can briefly enter the high amplitude branch before falling out of resonance during the downward sweep in (d). The inset shows a schematic amplitude dependence of the mechanical oscillation. Note that the detector saturates in the high-amplitude regimes in frames (c) and (d).
	\label{fig_duffing_vs_gate}}
\end{figure}

Additional results on the presence of the \(5^{\rm th}\) order nonlinearity are presented in Figure~\ref{fig_duffing_vs_gate}. This data set was measured at another charge degeneracy point located at \(V_{\rm G0}\approx10.5\) V during a separate cool down. Figure~\ref{fig_duffing_vs_gate} (a) shows a map of the mechanical oscillation induced change in \(I_{\rm SD}\) measured as a function of \(f_{\rm D}\) while the gate voltage is stepped across \(\alpha=0\) point located on the right side of the degeneracy point. The positions of the cross sectional resonance response curves in (b)-(e) are highlighted with vertical red lines in (a). Upward sweeps of the drive frequency are denoted in blue and downward sweeps in red.

At the lowest end of the presented gate voltage range, single-electron-tunneling-induced \(3^{\rm rd}\) order term is positive and dominates the high amplitude mechanical motion. Therefore, the resonance peak tilts to higher frequencies and results in Duffing type resonance response shown in Figure~\ref{fig_duffing_vs_gate} (b). At the upper end of the presented \(V_{\rm G0}\) range, electron-tunneling-induced \(3^{\rm rd}\) order term has turned negative and the resonance response peak tilts to lower frequencies resulting in hysteretic Duffing response shown in (e).

Figure~\ref{fig_duffing_vs_gate} (c) plots the frequency response at an intermediate \(V_{\rm G0}\) point, where the resonance peak is tilted to higher frequencies at high oscillation amplitudes due to \(5^{\rm th}\) order nonlinearity. The competing negative \(3^{\rm rd}\) order term is seen as continuous rise of the signal on the downward sweep, before the resonator jumps to the high amplitude branch. This bias point is similar to the one characterized in detail in Figure~\ref{fig:nlinresp}.

Interestingly, when the \(3^{\rm rd}\) order term is further decreased from the situation in Figure~\ref{fig_duffing_vs_gate} (c) by increasing \(V_{\rm G0}\), the resonance peak tilts first to lower frequencies at small oscillation amplitudes and then to higher frequencies at large oscillation amplitudes. This situation is presented in (d). Forward tilted branch is briefly probed during the downward frequency sweep when the measured current jumps from the continuous slope to a higher value, before collapsing down. Note that the detector is saturated at the highest amplitudes in Figures~\ref{fig_duffing_vs_gate} (c) and (d). The inset shows a schematic shape of the corresponding response peak for oscillation amplitude.

\clearpage
%===========================================================================================================================================
\section*{S3 Charge sensitivity}
Charge sensitivity \(\delta Q\) is defined as the charge noise \(\Delta Q\) of the system per measurement bandwidth \(B\)
\[
\delta Q=\Delta Q/\sqrt{B}.
\]
For a SET, this charge noise induces current or voltage noise of
\[
\sqrt{S_{A}}=\delta Q\left|\frac{\partial A_{\rm{SD}}}{\partial Q}\right|.
\]
where \(S_{A}\) is either the current- or voltage-noise spectral density and \(A_{\rm{SD}}\) is the source-drain current or voltage, respectively. For a given dB signal-to-noise ratio SNR this becomes
\[
\delta Q = \frac{\Delta q_{\rm rms}}{\sqrt{B}\cdot10^{\rm{SNR}/20}}.
\]
Here \(\Delta q_{\rm rms}\) is the root mean square gate charge variation caused by a known voltage signal on the gate.

In the main text we found that optimal charge sensitivity is reached by biasing the sample to a point where the regular Duffing parameter cancels out and the fifth order nonlinearity dominates high amplitude mechanical motion providing large \(\partial G/\partial\left(\delta f\right)\) and thus enhanced \(dG/dV_{\rm G}=\partial G/\partial V_{\rm G}-\partial G/\partial\left(\delta f\right)\times\partial f_{0}/\partial V_{\rm G}\). In general, for resonance peaks having two continuous slopes, the steeper instability-enhanced side provides larger signal due to the enhanced transducer gain. However, additional noise was found to be generated along with instability enhancement and thus larger signal did not lead to improved \(\delta Q\).

If island charge noise would be dominated by electrical noise on the gate electrode or capacitively coupled background charge fluctuation, the signal-to-noise ratio and thus the charge sensitivity would not depend on transducer gain. This is not the case since the charge sensitivity improves by a factor of \(1/2\) at 10 Hz and \(1/6\) at 1273 Hz. Figure 5 of the main text plots measured noise floors near \(\alpha\approx0\) with and without rf-excitation. In both cases the noise characteristics can be well fitted by a sum of two power spectral densities of two-level fluctuators and a white noise contribution
\[
S_{V}(f)=\sum_{i=1}^{2}\frac{A_{i}}{1+4\pi^2\tau_{i}^{2}f^{2}}+S(0).
\]
Here the fitting parameters are amplitude \(A_{i}\) and lifetime \(\tau_{i}\). The red curve with rf-excitation is reproduced using TLFs having lifetimes \(\tau_{1}=700\) ms and \(\tau_{2}=1\) ms. The blue curve without rf-drive is also reproduced with TLFs having the same lifetimes, \(\tau_{1}\) and \(\tau_{2}\), as the driven case. The magnitude of the plateau appearing at the highest frequencies is explained well by white noise contributions arising from the shot noise and the input noise of the voltage amplifier . The amplitudes \(A_{i}\) with and without drive are different and do not change by the same factor between these two cases implying that these TLFs couple to the charge noise in distinct manner. For an ordinary SET, the main noise sources in this frequency range are charge trapping centers, which may be located either in the substrate, at close vicinity to the island, or in the tunnel barriers. In addition, fluctuations in the tunnel resistances contribute to the \(1/f^{\alpha}\)-noise.

\begin{figure}[htb]
\begin{center}
      \includegraphics[]{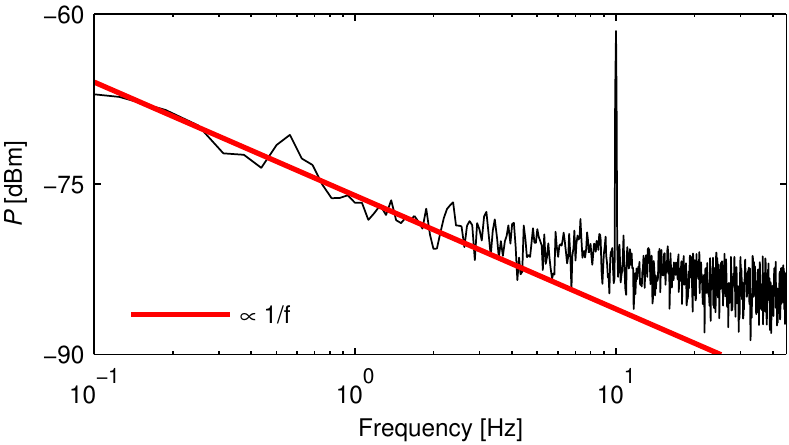}
    \caption{Noise spectrum measured under voltage bias at the maximum 	slope of conductance on the side of the Coulomb peak in Figure 2(b). We find sensitivity of \(\delta Q=7.6\) \textmu\(e/\sqrt{\mathrm{Hz}}\) for the conventional operation with \(V_{\rm SD}=0.25\) mV. The noise is calculated as the average current noise power in the vicinity of the 10~Hz peak. The \(1/f\)-noise-slope (red line) shows that the noise floor here starts to deviate from \(1/f\) already at a few Hz.
		}
		\label{fig_noise_spectra_coulomb_effect}
\end{center}
\end{figure}

\clearpage

\begin{figure}[htb]
\centering
  \begin{tabular}{cc}
    \includegraphics[angle=90]{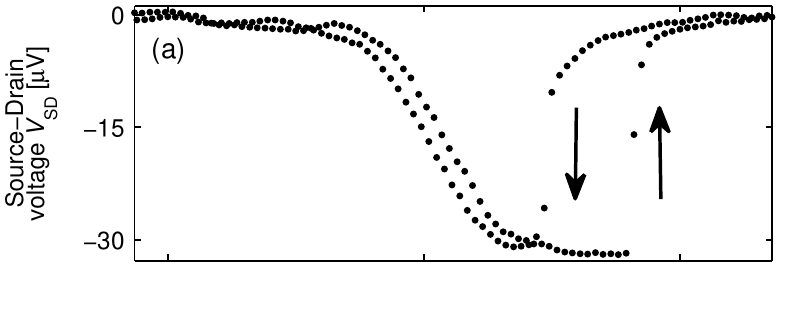}\\[-3.5ex]
    \includegraphics[]{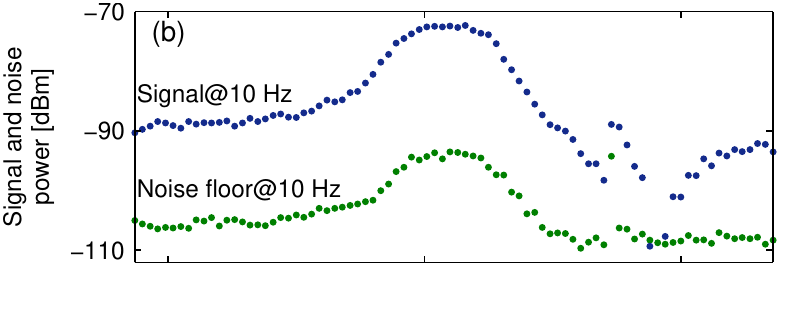}\\[-3ex]
    \includegraphics[]{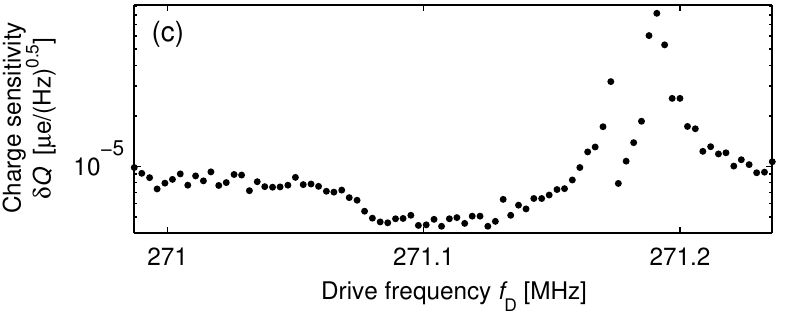}
  \end{tabular}
  \caption{Measurement of Figure 4 redone at \(f_{\rm R}=10\) Hz. Enhancement of the charge sensitivity is more modest than at
    1273 Hz, \(1/2\) opposed to \(1/6\).
		}
\label{fig_sensitivity_cbias_10hz}
\end{figure}

%\end{suppinfo}
%\bibliography{references}

%\end{document}